\documentclass[pra,twocolumn,superscriptaddress]{revtex4-1}

\usepackage{graphicx,color}
\usepackage{amsmath}
\usepackage{amssymb}
\usepackage{hyperref}
\usepackage[normalem]{ulem}
\usepackage{dsfont}
\usepackage{textcomp}

\newcommand{\ket}[1]{|{#1}\rangle}
\newcommand{\bra}[1]{\langle{#1}|}

 \newcommand\beq            {\begin{equation}}
 \newcommand\eeq           {\end{equation}}
 \newcommand\bwt         {\begin{widetext}}
 \newcommand\ewt         {\end{widetext}}

\begin{document}

\title{Reaching the quantum Hall regime with rotating Rydberg-dressed atoms}

\author{Michele Burrello}
\email{for correspondence: michele.burrello@nbi.ku.dk}
\affiliation{Niels Bohr International Academy and Center for Quantum Devices, University of Copenhagen, Lyngbyvej 2,
2100 Copenhagen, Denmark}

\author{Igor Lesanovsky}
\affiliation{Institut f\"ur Theoretische Physik, Universit\"at T\"ubingen, Auf der Morgenstelle 14, 72076 T\"ubingen, Germany}
\affiliation{School of Physics and Astronomy and Centre for the Mathematics and Theoretical Physics of Quantum Non-Equilibrium Systems,The University of Nottingham, Nottingham, NG7 2RD, United Kingdom}

\author{Andrea Trombettoni}
\affiliation{Department of Physics, University of Trieste, Strada Costiera 11, I-34151 Trieste, Italy}
\affiliation{CNR-IOM DEMOCRITOS Simulation Center, via Bonomea 265, I-34136 Trieste, Italy.}
\affiliation{SISSA and INFN, Sezione di Trieste,
via Bonomea 265, I-34136 Trieste, Italy.}

\begin{abstract}
Despite the striking progress in the field of quantum gases, one of their much anticipated applications -- the simulation
  of quantum Hall states -- remains elusive: all experimental approaches so far have failed in reaching a sufficiently small ratio between atom and vortex densities. In this paper we consider
  rotating Rydberg--dressed atoms in magnetic traps: these gases offer strong and tunable non-local repulsive interactions and very low densities; hence they provide an exceptional platform to reach the quantum Hall regime. 
	Based on the Lindemann criterion and the analysis of the interplay of the length scales of the system, we show that there exists an optimal value of the dressing parameters that minimizes the ratio between the filling factor of the system and its critical value to enter the Hall regime, thus making it possible to reach this strongly--correlated phase for more than 1000 atoms under realistic conditions.
\end{abstract}

\maketitle

\textit{Introduction.-} In the last decades ultracold atoms allowed for the study and quantum simulation
of a plethora of quantum many-body effects \cite{bloch08}.
Despite the impressive successes, however, one of the most anticipated applications, so far, has resisted many attempts of implementation: reaching the quantum Hall (QH) regime. 

Since the realization of Bose--Einstein condensates in the mid $90$s \cite{dalfovo99}, the nucleation of quantized vortices in rotating ultracold atoms
\cite{cornell04,dalibard04,fetter09} naturally suggested the possibility of creating QH states by rotating
strongly interacting gases. The dynamics of atomic clouds in the rotating frame can indeed be described
in terms of Coriolis/Lorentz forces, which define in turn the appearance of a 
synthetic magnetic field $B$ for neutral atoms \cite{baym05,cooper08}.

Reaching the QH regime, however, requires strong magnetic fields: it is necessary
to achieve angular velocities extremely close 
to the critical value set by the trapping potentials -- so close that, for practical purposes, 
this possibility was experimentally ruled out.

Alternative approaches based on optically induced gauge potentials have been proposed and tested \cite{spielman09,spielman13,dalibard11,goldman13rev}, but, also in this case, the simulated magnetic fields were not strong enough to access the QH regime.

In all these experiments, the interactions among the atoms were effectively contact interactions. 
In the last years, however, atoms with long--range interactions have been at the focus of intensive investigations,  in the cases of both dipolar gases \cite{modugno2019,pfau2019,ferlaino2019,ferlaino2019b} and Rydberg--dressed atoms with strong van der Waals interactions \cite{gross15,gross16,bernien17}. Intuitively, such strong long--range repulsions favor the formation of gases with lower densities, thus making it easier to achieve the low filling factors required for QH states.

In this work, we consider ultracold bosonic gases subject to long--range repulsive interactions and synthetic gauge fields.
We will show that moderate van der Waals interactions help in reaching the ratio between atomic and vortex densities required for the onset of the QH regime. We will focus on Rydberg-dressed atoms, which allows us to tune the effective value of the interactions, and we will mostly address the case of synthetic fields obtained by rotation, since for a realistic number of atoms and other parameters this technique provides better results than optically generated magnetic fields.

Our main result is that the long-range interaction between Rydberg-dressed atoms facilitates reaching lower filling factors in comparison with ground state atoms subject to the same artificial magnetic field. In particular, we show that the melting transition of the superfluid vortex lattice is favored by this interaction, and we hypothesize that this signals the onset of the QH regime, shown to appear for small filling factors in recent works \cite{Grusdt2013,Grass2018}.

\section{The main idea} 

When a Bose--Einstein condensate is rotating and its angular velocity, and thus the 
artificial magnetic field, is progressively increased, 
vortices enter the superfluid and arrange themselves in denser and denser 
triangular lattices. For strong magnetic fields and confinement in the direction of the rotation axis, the condensate enters the so--called lowest Landau level (LLL) regime, in which the vortex size scales with the magnetic length $l_{B}=\sqrt{\hbar/B}$ \cite{noteb}. The further transition from the LLL to the QH regime corresponds to the melting of the vortex lattice into a strongly--correlated phase 
\cite{baym05,cooper08,rozhkov96}, driven by the quantum fluctuations 
of the vortices. The critical value of $B$ for this transition can be estimated from the Lindemann criterion: the lattice melts when the ratio $l_L / l_B$ between the quantum fluctuation $l_L$ of the positions of the vortex cores and the intervortex distance (proportional to $l_B$) reaches a critical value $\alpha_L$. The main parameter to characterize these systems is the filling factor 
\begin{equation}
\nu={n}/{n_v}={2 \pi \hbar n}/{B},
\label{ratio}
\end{equation} 
given by the ratio between the atom and the vortex areal densities $n$ and 
$n_v$, respectively. $\nu$ is proportional to $(l_B/l_L)^2$ \cite{cooper08,rozhkov96}, such that the gas enters the QH regime for $\nu$ smaller than a critical value $\nu_c$. For weakly interacting bosons, the most conservative estimates give a critical filling factor $\nu_{c,0} \lesssim 6$ \cite{fetter09,cooper08,sinova02,cooper01,regnault2004}. Here the subscript $0$ specifies that this is the critical value for weak contact interactions. We will show below that such a critical value does not hold for interactions beyond a certain threshold.

The limit $\nu=\nu_{c,0}$ is extremely difficult to reach for rotating gases: the centrifugal limit of the in--plane trapping poses a severe bound on the maximal angular velocity, and thus on the maximal field $B$; on the experimental side, the smallest parameters 
$\nu$ achieved \cite{cornell04,dalibard04} are about $\nu\sim 300$. Much smaller filling factors ($\nu \approx 1$) were instead reached in rotating optical microtraps, but only for a very small number of atoms, $N\approx 5$ \cite{gemelke10}.

As is evident from \eqref{ratio}, besides increasing $B$, there is another strategy to lower $\nu$: namely,
to reduce the two-dimensional (2D) atom density $n$. To this purpose, we consider Rydberg--dressed atoms: we will first estimate the behavior of the superfluid density as a function of the interactions and we will discuss its implications for the phase diagram of these gases. We will show that, for moderate long--range interactions, the range of parameters for which the QH regime exists is enhanced, whereas, when the interactions exceed a certain threshold, it is suppressed. Hence there exists an optimal value for the Rydberg dressing that minimizes the magnetic field needed to enter the QH regime.

\section{The physical system} 

A rotating atomic cloud has a dynamics equivalent to charged particles in a magnetic field, as an effect of the Coriolis force \cite{fetter09,baym05,cooper08}. We consider a Bose--Einstein condensate subject to a harmonic potential with trapping frequency $\Omega_{\rm tr}$, and rotating with frequency $\Omega_{\rm rot}$. As discussed in App. \ref{app:rotating}, in the rotating frame the single--particle Hamiltonian is 
\begin{equation} \label{hameff}
H_{\rm rot} =  \frac{\left(-i\hbar\vec{\nabla} + \vec{A}\right)^2}{2m} +\frac{m\left(\Omega_{\rm tr}^2-\Omega_{\rm rot}^2\right) r^2}{2}\,.
\end{equation} 
Here we have introduced the vector potential $\vec{A}= m\Omega_{\rm rot} (y,-x,0)$ and $m$ is the atomic mass. The resulting artificial magnetic field lies along $\hat{z}$ with intensity $B\equiv 2m\Omega_{\rm rot}$, and the effective in--plane trapping potential is reduced by the centrifugal force. In particular, we define the ratio $\gamma= \Omega_{\rm rot} / \Omega_{\rm tr}<1$, such that the effective trapping frequency is $\sqrt{1-\gamma^2} \Omega_{\rm tr}$.

The Rydberg dressing amounts to a weak coupling between a ground state $\ket{g}$ of the chosen atoms and a Rydberg state $\ket{e}$. This coupling is obtained through laser beams propagating 
in the $\hat{z}$ direction such that the 
related Rabi frequency $\Omega=|\Omega|e^{i\phi_\Omega}$ does not depend 
on the position in the $xy$ plane. We consider an effective detuning for this coupling $2\delta \gg \Omega$ such that, for a single atom, its lowest energy state becomes
\begin{equation}
\ket{\tilde{g}}=-e^{i\phi_\Omega}\sin\left(\theta/2\right) \ket{e} +  \cos(\theta/2) \ket{g}\,, \\
\end{equation}
where $\tan\theta = {|\Omega|}/\delta$ (see App. \ref{app:rotating}). A generic Rydberg interaction $H_{\rm int}= V(\vec{r_1} - \vec{r_2}) \ket{ee}\bra{ee}$ results in a typical long--range interaction that decays like $\sin^4(\theta/2)V$ for large separations, and is characterized by a plateau $2\delta\sin^4(\theta/2)$ for short distances
\begin{equation} \label{Rydpot}
V_{\rm Rydberg} (\left|\vec{r_1}-\vec{r_2}\right|) \approx \frac{C_6\sin^4(\theta/2)}{a^6 + \left|\vec{r_1}-\vec{r_2}\right|^6}\,,
\end{equation}
where we introduced the Rydberg radius $a\approx \left(C_6/2\delta\right)^{1/6}$ and the van der Waals coefficient $C_6$. For the typical Rydberg state $43S_{1/2}$ of $^{87}$Rb, the interaction is given by $C_6 \approx h\cdot 2.4$ GHz \textmu m$^6$ \cite{low2012} with $a \approx 2.0$ \textmu m for a mixing angle $\theta=0.05$.

\section{Density of a 2D Rydberg-dressed gas}

The effect of the strong interactions of Rydberg-dressed atoms on their density can be estimated with a variational Gross--Pitaevskii calculation. In particular, we consider the isotropic interaction in Eq. \eqref{Rydpot} and we focus on gases in the lowest Landau level (LLL) regime with strong confinement in the third direction, such that a 2D approximation holds.
We combine a description of the van der Waals interactions in the spirit of \cite{lewenstein2000} and the variational ansatz introduced in \cite{aftalion2005} for the superfluid vortex lattice, defined by the many-body wavefunction  
$\psi_s(\vec{r})=p(\vec{r})e^{-\frac{r^2}{2s^2}}\sqrt{N/\pi s^2}$
where $r=|\vec{r}|$. This wavefunction is characterized by a periodic modulation $p(\vec{r})$, which defines the triangular vortex lattice, and a slow-varying Gaussian envelope of width $s$. The average density of the system is approximately set only by its long--distance behavior, thus by the Gaussian envelope.  It is therefore a function of the variational parameter $s$, which is linear in the average distance from the center. By averaging over the modulation (see App. \ref{app:rotating}), the mean-field energy results in 
\begin{multline} \label{GP}
E
\approx \frac{\hbar^2 N}{2ms^2} + \frac{Nm\Omega_{\rm tr}^2(1-\gamma^2)s^2}{2} +  \\
\frac{N^2}{s^2}\left[\frac{bg}{4\pi}+\int_0^\infty rdr e^{-\frac{r^2}{2s^2} } \frac{V_6}{a^6 + r^6} \right] \,.
\end{multline}
Here $g$ is the 2D contact interaction parameter and $V_6 = \sin^4(\theta/2) C_6$ is the effective van der Waals coupling constant. The numerical factor $b \approx 1.1596$ effectively increases the contact interaction in the LLL approximation due to the inhomogeneity introduced by the triangular vortex lattice \cite{aftalion2005} (see App. \ref{app:rotating}).

The potential energy and the long--range interaction are estimated by separating their rapidly and slowly oscillating contributions with a procedure analogous to the so-called averaged vortex approximation \cite{ho2001} for $s \gg l_B$: within each unit cell of the lattice, the value of the Gaussian factor, harmonic potential and van der Waals interaction is considered approximately constant, such that the modulation averages to 1 and does not affect the final result.
 
By expanding the integral in Eq. \eqref{GP} in series of $s^{-1}$ for $s\gg a$ and minimizing the energy, we find
\begin{align} \label{effint}
&s= \sqrt[4]{\frac{Nmg' +2\pi\hbar^2}{2\pi m^2\Omega_{\rm tr}^2(1-\gamma^2)}} \,, \\
&g' = bg +\frac{4\pi^2 V_6}{3\sqrt{3}a^4} \approx 1.16g +7.6\,\frac{V_6}{a^4}\,.
\end{align}
With the introduction of the Rydberg dressing, the contact interaction $g$ must be effectively replaced by the considerably stronger interaction amplitude $g'$, derived by the large-$s$ expansion of the integral in Eq. \eqref{GP}. In the following, we use Eq. \eqref{effint} to estimate the gas density. We adopt in particular $n=N/(4\pi s^2)$: $n$ scales like $1/\sqrt{g'}$ when the kinetic energy is negligible.

For a gas of $^{87}$Rb, the typical strength of the contact interaction is $g\approx h \cdot 23$ Hz \textmu m$^2$ (for a trapping frequency along $\hat{z}$ given by $\Omega_z = 2\pi$ kHz). For the Rydberg state $43S_{1/2}$ at mixing angle $\theta=0.05$, $V_6/a^4 \approx h \cdot 61 {\rm Hz}$ \textmu m$^2$. The ratio between the contact and Rydberg interactions is thus  $(7.6 V_6/a^4)/bg   \approx 18$, which implies that $n$ and $\nu$ decrease by a factor ${\sf f}_\nu \sim 4.5$.

\begin{figure}[t]
\includegraphics[width=0.85\columnwidth]{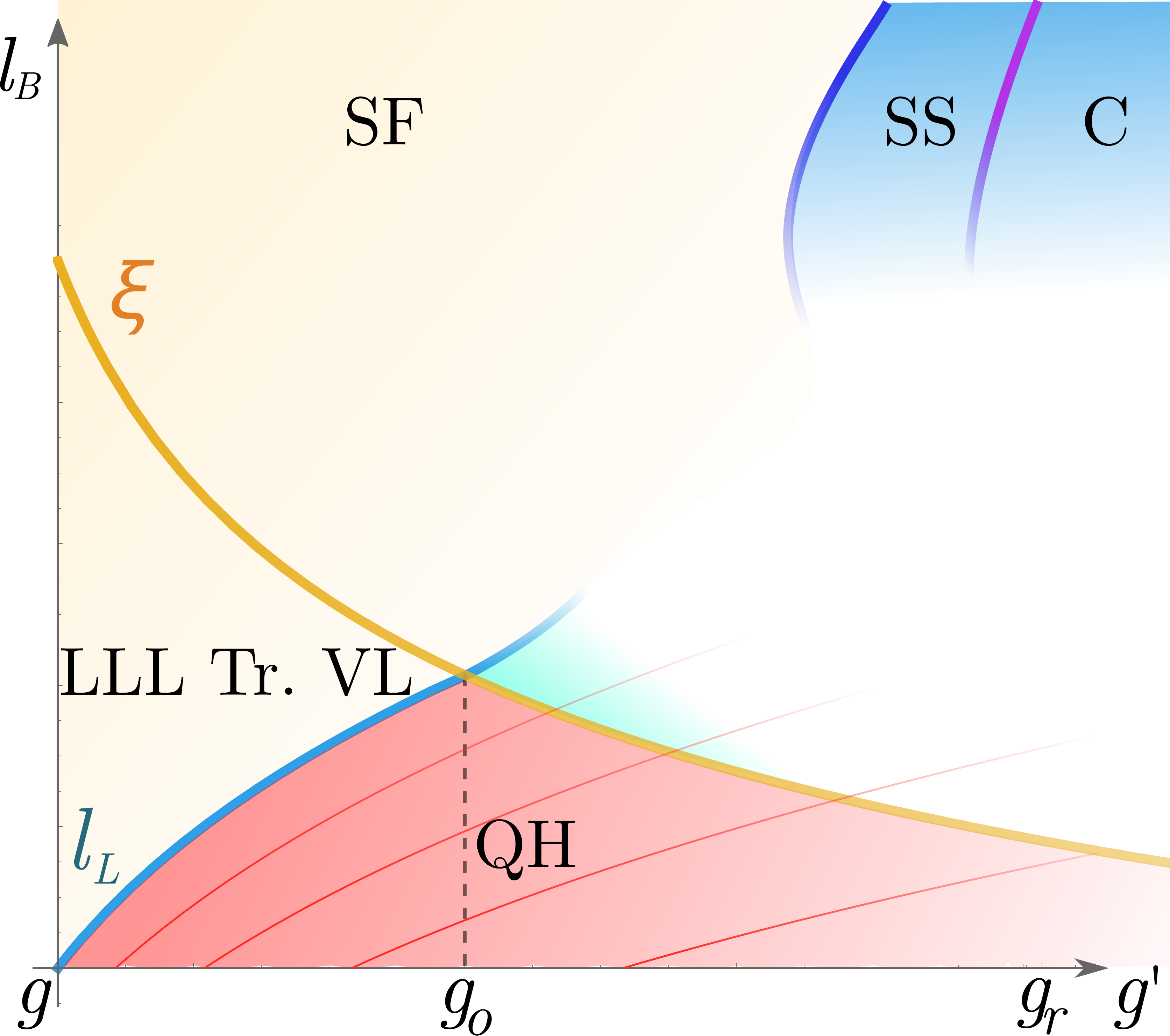}
\caption{Qualitative phase diagram of the Rydberg dressed gas as a function of the magnetic length $l_B$ and the interaction parameter $g'$. For $l_B \to \infty$ (thus $B\to 0$), the system displays a superfluid (SF), supersolid (SS) or crystal (C) state. For interactions $g'<g_o$, by decreasing $l_B$, the system presents first a cross--over into the lowest Landau level triangular vortex lattice (LLL Tr. VL) (at $l_B \sim \xi/\alpha_\xi$), then a transition to the QH regime for $l_B \sim l_L/\alpha_L$. For $g'>g_o$, instead, the LLL vortex lattice phase disappears and the QH regime is reached for lower values of the filling factor. The thin red lines are lines at constant filling factor.} \label{fig:pd}
\end{figure}

\section{Low--density Rydberg--dressed gases} 

To understand the effect of the long--range interaction on the onset of the QH regime, let us analyze the main changes in the phase diagram of the rotating condensate (Fig. \ref{fig:pd}).
For weak or no Rydberg dressing, the phase diagram can be intuitively understood from the comparison of three distinct length scales \cite{baym05,cooper08,cooper01}: the magnetic length $l_B$, the superfluid healing length $\xi =\hbar / \sqrt{2mng'} \propto (g')^{-1/4}$ and the Lindemann length \cite{cooper08,rozhkov96} $l_L \approx \sqrt{1/\pi n} \propto (g')^{1/4}$. By increasing $B$, the system evolves from the pure superfluid phase with $l_L/\alpha_L < \xi/\alpha_\xi < l_B$, to the vortex lattice phase in the LLL regime with $l_L/\alpha_L < l_B < \xi/\alpha_\xi$, to the QH phase where $l_B<l_L/\alpha_L < \xi/\alpha_\xi$ (left side of Fig. \ref{fig:pd}). Here $\alpha_\xi \approx 0.3$ \cite{cornell04,coddington04,baym04} is the ratio $\xi/l_B$ at the crossover to the LLL regime \cite{baymnote}, whereas $\alpha_L \approx 0.4$ is the Lindemann parameter \cite{cooper08,rozhkov96}, corrected by the geometrical factor for triangular lattices. In particular, the relation $l_L/\alpha_L = l_B$ provides the estimate $\nu_{c,0} \approx 14$ which, however, must be corrected to account for collective modes of the vortices \cite{sinova02}, resulting in $\nu_{c,0} \approx 8$ \cite{cooper08}. Even lower values, $\nu_{c,0}\lesssim 6$, are suggested by numerical works \cite{cooper01,regnault2004}; therefore, we introduce an effective Lindemann factor $\alpha_L'=\sqrt{2/\nu_{c,0}}$.

\begin{figure}
\includegraphics[width=0.95\columnwidth]{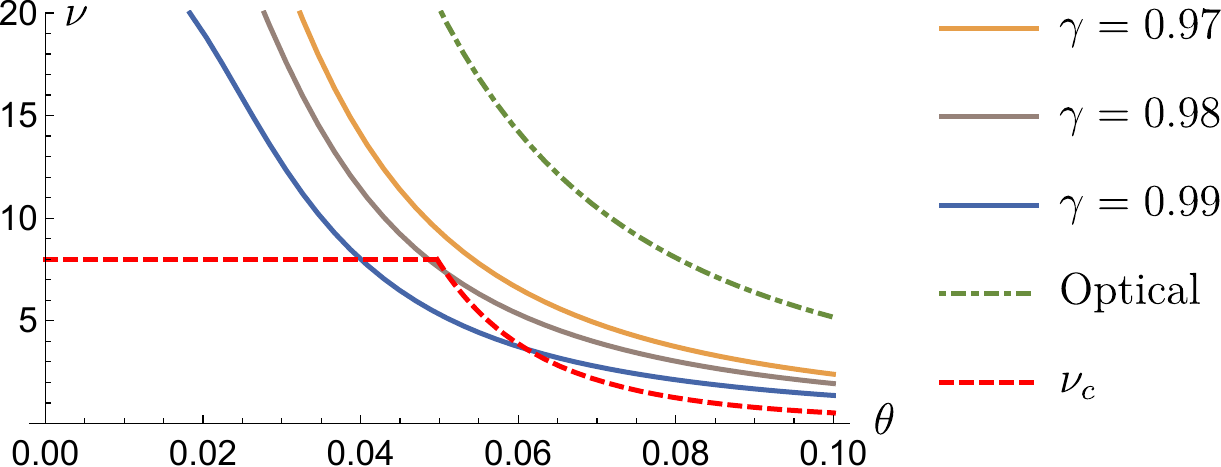}
\caption{Filling factor as a function of the mixing angle $\theta$ for $\delta=20$MHz and $N=15000$. The solid lines correspond to rotating gases with $\Omega_{\rm tr}=2\pi 100$Hz and different values of $\gamma$. The green dot--dashed line is an estimate of $\nu$ for optically induced magnetic fields based on \cite{juze06} obtained with counterpropagating Gaussian beams (waist $w=50$\textmu m, wavelength $\lambda=790$nm) for $\Omega_{\rm tr}=2\pi 20$Hz. The red dashed line represents the critical filling factor assuming $\nu_{c,0}=8$; its cusp determines the optimal interaction point at $\theta \approx 0.05$.} \label{fig:nu_theta}
\end{figure}

The Rydberg interaction modifies this scenario because it decreases the ratio $\xi/l_L$ by the factor ${\sf f}_\nu$. Hence, for sufficiently large mixing angles, the LLL vortex lattice phase is suppressed and additional supersolid phases may appear \cite{Pohl2012}. Therefore, for $\xi /\alpha_{\xi} < l_L/\alpha_L'$ the usual Lindemann criterion cannot be applied for the onset of the QH phase. A new estimate of $\nu_c$, though, can be obtained by imposing $l_B < \xi / \alpha_\xi$, which results in 
\begin{equation} \label{nuc2}
\nu_c  = \frac{\pi \hbar^2}{ m g' \alpha_\xi^2}\,, \quad \text{for } \xi /\alpha_{\xi} < l_L/\alpha_L'\,.
\end{equation}
Globally, the ratio $\nu/\nu_c$ can be minimized by interactions such that $\xi /\alpha_{\xi} \approx l_L/\alpha_L'$, hence for an optimal value $g_o$ of the parameter $g'$ given by:
 \begin{equation} 
g_o  \equiv \frac{\pi \hbar^2}{m \nu_{c,0} \alpha_\xi^2}\,.
\end{equation}
$g_o$ ranges in $h \cdot 280 - 640\, {\rm Hz}$ \textmu m$^2$ for $\nu_{c,o} = 14 - 6$. For the van der Waals interactions of the $^{87}$Rb state 43$S$, the condition $\xi /\alpha_\xi \approx l_L/\alpha_L'$ is met for mixing angles $\theta \approx 0.04 - 0.06$ (see Fig. \ref{fig:nu_theta}). For $g'=g_0$ and $B$ such that $\nu=\nu_{c,0}$, the system lies at a critical point that separates four different phases (see Fig. \ref{fig:pd}): the triangular vortex lattice in the LLL regime, appearing for $g'<g_o$ $B$ constant; the QH regime obtained for $g'=g_o$ by increasing $B$; the superfluid vortex lattice for smaller values of $B$ and a strongly interacting phase for $g'>g_o$.

The interaction amplitude $g_o$ lies at the edge of the regime of validity of the mean--field energy estimate: the effective scattering length results $a_s \approx 0.83$ \textmu m for $\theta=0.05$ (see App. \ref{app:scatter}), to be compared with the average interatom distance of about $0.7$ \textmu m for $\gamma=0.98$ and $\Omega_{\rm tr}=2\pi$ 100 Hz. The gas, in proximity to $g_o$, reaches a regime that cannot be considered any longer ultradilute and the breakdown of the mean--field approximation signals indeed that the superfluid approached an unstable point.

We also observe that for the parameters adopted in Fig. \ref{fig:nu_theta}, in proximity to the critical point at $\theta=0.05$, the ratio of the first two Haldane pseudopotential \cite{Haldane,Grusdt2013} results $\mathcal{V}_2/\mathcal{V}_0 \approx 0.22$, close to the critical value $0.20$ identified in Ref. \cite{cooper2005} for the transition between a triangular and a square vortex lattice \cite{zhang2005}. This suggests the onset of several phases with broken translational invariance for $g'>g_o$ and it is consistent with the behavior of the system at both strong and weak magnetic fields.

Concerning strong magnetic fields and interactions ($g'>g_o$), the healing length decreases and, with it, the extent of the QH phase. For intermediate filling factors, this gives rise to several inhomogeneous phases including stripe and bubble states \cite{cooper2005}.  In the extreme regime with low filling factors ($\nu \lesssim 1$), it is known that a competition between QH states and Wigner crystals appears \cite{Grusdt2013,Grass2018}. 

Concerning weak magnetic fields (upper part of Fig. \ref{fig:pd}), the system is no longer in the LLL regime and different mean-field ans\"atze are required. The strong Rydberg interactions cause a spontaneous breaking of translational symmetry, corresponding to supersolid and crystal phases \cite{Henckel2010,igor10,Pohl2012,Macri2014}.
Mean--field analyses \cite{Henckel2010,Macri2013,Macri2014} estimate the onset of the crystal phase based on the dispersion of the superfluid roton. The roton gap closes, which signals an instability towards a crystalline phase, when the interaction energy density $u = 3\sqrt{3} m n a^2  g' / h^2$ reaches a critical value $u_c \gtrsim 40$, corresponding to $g' > g_r \approx 1600\cdot 8\pi h^4 / (27 m^4 a^4 N \Omega_{\rm tr}^2) \approx h \cdot 2.5 \rm{kHz}$ \textmu m$^2$ (for 15000 particles with $\Omega_{\rm tr} = 2\pi\cdot 100$Hz). Additionally, in our regime of interest $(a^2 n >1)$, a supersolid phase is expected for intermediate interactions with $30\lesssim u \lesssim 40$ \cite{Macri2014} (see also the recent experiments \cite{pfau2019,modugno2019,ferlaino2019,ferlaino2019b} in elongated dipolar clouds). By increasing $B$, the value of the critical interaction is non--monotonic \cite{Pohl2012,Sinha2005} and it is hard to extrapolate the behavior of the system for $g'>g_o$: a probable scenario is that several phases with broken translational symmetry alternate for large interactions. 

Our estimates for the melting point of the LLL vortex lattice relies on the Gross-Pitaevskii approximation of the gas density. We emphasize that the validity of these mean-field estimates of the density and energy of superfluid systems in proximity to the breaking of superfluidity has been successfully used in many different cases. An example is given by superfluid-Mott phase transitions, where the Gross--Pitaevskii equation provides reasonable results for the energy and Hamiltonian parameters also in the presence of strong quantum fluctuations \cite{giovanazzi09}. In the case of long--range interactions, these mean--field analyses give a fair estimate of the superfluid breakdown and even a reasonable estimate of the energy beyond the transition point from superfluid to supersolid \cite{Macri2013}. Therefore we expect Eq. \eqref{effint} to capture the main physical features of the system also in proximity to the LLL vortex lattice melting.

\section{ The optimal filling factor} 

By comparing the density of atoms obtained from Eq. \eqref{effint} and the artificial magnetic field, the filling factor (Eq. \eqref{ratio}) results:
\begin{equation} \label{nurot}
\nu = \frac{N}{4\gamma} \sqrt{\frac{2\pi(1-\gamma^2)\hbar^2}{Nmg'+2\pi\hbar^2}}\,.
\end{equation}
From Eq. \eqref{nuc2} and \eqref{nurot} we derive its optimal value
\begin{equation} \label{nuo}
\nu_o = \frac{N}{2\gamma}\sqrt{\frac{(1-\gamma^2)\alpha_\xi^2}{N\alpha_L^{\prime 2}+4\alpha_\xi^2}} = \frac{N}{2\gamma}\sqrt{\frac{(1-\gamma^2)\alpha_\xi^2}{2N/\nu_{c,0}+4\alpha_\xi^2}}\,.
\end{equation}

\begin{figure}[t]
\begin{center}
\includegraphics[width=0.95\columnwidth]{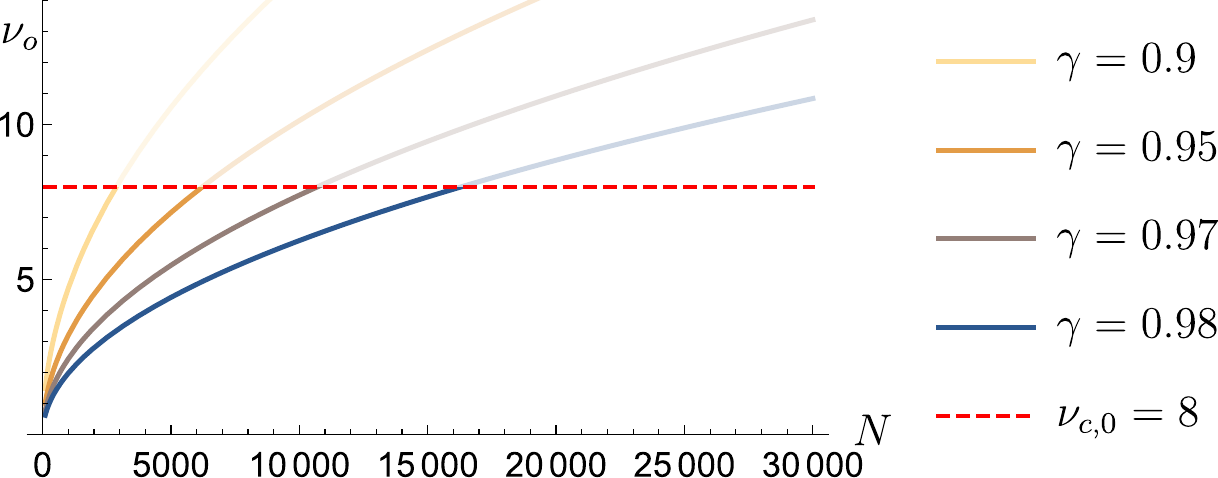}\llap{
  \parbox[b]{14.5cm}{(a)\\\rule{0ex}{3.25cm}}}\\
\vspace{0.2cm}
	
\includegraphics[width=0.95\columnwidth]{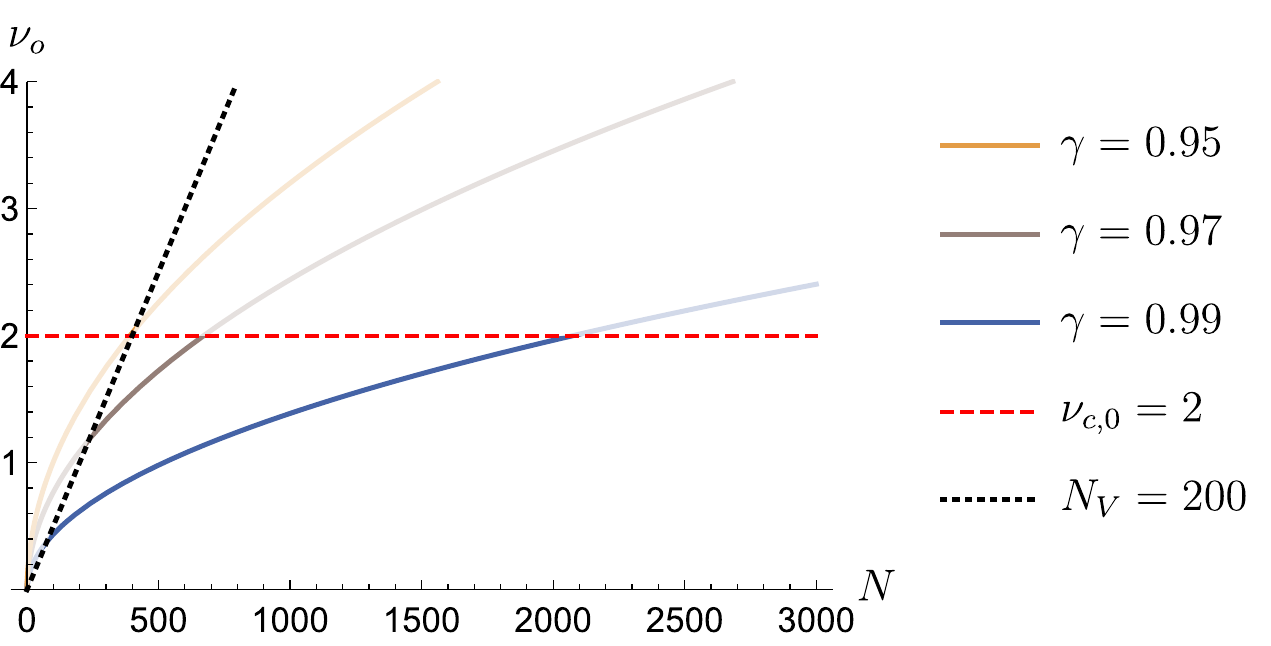}\llap{
  \parbox[b]{14.5cm}{(b)\\\rule{0ex}{3.25cm}}}
\end{center}
\caption{Filling factor at the optimal interaction point (Eq. \eqref{nuo}) as a function of the atom number $N$ assuming $\nu_{c,0}=8$ [panel (a)] and $\nu_{c,0}=2$ [panel(b)]. The critical filling is depicted by dashed lines and the system reaches the QH regime for $\nu_o<\nu_{c,0}$. For $\nu_{c,0}=8$, the Rydberg-dressed gas reaches the QH regime for a large range of values of $N$ for rotation frequencies with $\gamma > 0.95$. For $\nu_{0,c}=2$ the QH regime is reached for systems with $N\lesssim 2000$ at $\gamma=0.99$. In (b) the area on the right of the dotted black line corresponds to system with at least 200 magnetic fluxes.} \label{fig:nuN}
\end{figure}

In Fig. \ref{fig:nuN}(a) we display $\nu_o$ obtained from $\nu_{c,0}=8$  \cite{fetter09,cooper08,sinova02}. Under this assumption the QH regime is reached for $\gamma=0.95$ for clouds of $N \lesssim 5000$ atoms. Based on the previous interaction parameters, the optimal mixing angle corresponds to $\theta \approx 0.05$. Numerical studies \cite{cooper01,regnault2004} suggest that the critical filling factor can be smaller; it is therefore useful to consider also more restrictive values of $\nu_{c,0}$: in Fig. \ref{fig:nuN}(b) we present the values of $\nu_o$ for $\nu_{c,0}=2$. Since we aim at obtaining the QH regime for a mesoscopic gas, we included a constraint on the number $N_V$ of magnetic fluxes in the system: the dotted black line corresponds to the limit of 200 fluxes. From the plot we see that the QH regime is within reach for a gas of 500 atoms in a trap with $\gamma=0.97$ when the dressing is chosen close to the optimal point, at $\theta \approx 0.07$ for the considered Rydberg state 43$S$. Larger numbers of atoms are sustainable for larger $\gamma$.

To realize the considered rotating regime, the atoms must first be loaded in a magnetic trap; then, the rotation can be imparted through a deformation of the trap (see App. \ref{app:rotating}) and $N$ can be varied through evaporative techniques \cite{cornell04,cornell2003}. Finally the Rydberg dressing is switched on. The decay time of the Rydberg state 43$S$ at 300K is $\tau_{43}\approx 42$\textmu s \cite{low2012}, resulting in a decay time of the dressed state $\tau\approx \tau_{43}/\sin(\theta/2)^2 \approx 67$ms for $\theta=0.05$. This must be compared with the rotation period of about $10$ms. To increase the decay time $\tau$, more excited Rydberg states can be considered, resulting in a larger $\tau_n$ and a smaller mixing at the optimal point.

\section{Optically induced magnetic fields} 

Our estimate of the filling factor \eqref{nurot} is based on artificial magnetic fields obtained through rotation. Different approaches, based on fully optical setups, have also been successfully applied for bosonic gases \cite{spielman09,spielman13}, for example, exploiting position--dependent Raman couplings \cite{juze06}. These setups, however, are usually less convenient to reach small filling factors: $B$ is typically proportional to $(\lambda w)^{-1}$ with $\lambda \approx 790$nm being the Gaussian Raman lasers' wavelength and $w \sim 170$ \textmu m their waist \cite{spielman13}; therefore, to obtain a magnetic field approximately uniform over distances comparable with $s$, the typical value of $B$ is half of the value considered in the rotating case. Furthermore, the optical setups do not present the centrifugal reduction of the harmonic potential, such that the factor $\sqrt{1-\gamma^2}$ disappears from Eq. \eqref{nurot}, thus increasing the resulting $\nu$ for the same $\Omega_{\rm tr}$. In Fig. \ref{fig:nu_theta} we compare the rotating gases with a system with optically--induced fields for realistic parameters; for the optical realization we considered a lower trapping frequency to compensate for the missing centrifugal term.

\section{Conclusions} 

We have shown that a combination of Rydberg dressing and rotating traps can drastically reduce the filling factors $\nu$ obtained in rotating Rb gases. Even considering the worst--case scenario of a critical filling factor $\nu_{c,0}=2$, our estimates show that the quantum Hall regime can be reached for 2D gases of about 1000 atoms by introducing an optimal Rydberg dressing at $\theta \approx 0.07$. To increase the efficiency of this scheme, thus the number of atoms in the system, our proposal can be additionally combined with techniques for preparing many--body states with large angular momentum \cite{roncaglia11,ho16}.

\begin{acknowledgments} We warmly thank N. Cooper, L. Fallani, G. Juzeli\=unas, T. Macr\'i, C. Pethick and S. Simon for useful discussions. M. B. is supported by the Villum Foundation (Research Grant No. 25310). I. L. acknowledges support from EPSRC [Grant No. EP/N03404X/1] and from the ``Wissenschaftler--R\"uckkehrprogramm  GSO/CZS'' of  the Carl-Zeiss-Stiftung and the German Scholars Organization e.V..
\end{acknowledgments}

\onecolumngrid
\appendix

\section{Rydberg dressed atoms in a rotating frame and the Gross--Pitaevskii energy} \label{app:rotating}

The system we analyze relies on the combination of three elements: a quadratic trapping potential, a rotation-induced artificial gauge potential and the Rydberg dressing necessary to obtain strong repulsive interactions.

To realize such a system, we consider $^{87}$Rb atoms and we select a 5$S$ ground state $\ket{g} = \ket{5S_{1/2},F=2,m_F=2}$. As discussed in \cite{low2012,heidemann07}, such a state can be unambiguously dressed with a Rydberg excited states with the same $F$ and $m_F$ quantum numbers through a two-photon process via a state 5$P$. In this way, transitions to Rydberg states with $F'=1$ are forbidden due to the selection rules of the excitation scheme considered. In particular, for our numerical estimates, we considered the state $\ket{e}=\ket{43S_{1/2},F=2,m_F=2}$, but states with larger ${\sf n}$ can be considered as well.

The states $\ket{g}$ and $\ket{e}$ share all the angular quantum numbers and, consequently, they share the same magnetic moment. This makes it suitable to trap them with the same magnetic trap, following, for example, the techniques adopted in \cite{cornell04,cornell96,cornell2003} (see also \cite{bounds2018} for different trapping schemes for Sr atoms). We considered, in particular, a time--averaged orbiting potential with an effective frequency $\Omega_{\rm tr}=2\pi$ 100 Hz. A large angular momentum can be imparted to the cloud by elliptically deforming the magnetic trap in the horizontal plane and suddenly changing the angle of the deformation \cite{cornell2003}. The axial symmetry is then restored.
The effect of the rotation on the motion of the center of mass of the atoms is to introduce an effective vector potential  $\vec{A}=m\Omega_{\rm rot} (y,-x)$. 
In the following, we assume that the Rydberg dressing is created through lasers propagating along the rotation axis and centered with respect to the trap, such that they do not explicitly break rotational symmetry. These lasers can be turned on after the system is put in rotation and we remark that Doppler effects are negligible for realistic rotation frequencies.

In the rotating frame, the single-atom Hamiltonian reads
\begin{multline}
H_{\rm RF} = R(t) (H_{\rm kin} + \frac{m}{2}\Omega_{tr}^2 r^2 + H_{\rm dress}) R^\dag ( t) -i R (t) \partial_t R^\dag (t) = \\
 =\left[\frac{-\hbar^2{\vec{\nabla}}^2}{2m} -\Omega_{\rm rot} L_z +\frac{m\Omega_{\rm tr}^2 r^2}{2} \right] \mathbb{I} + \begin{pmatrix} E_R - J_r\Omega_{\rm rot}  & \Omega \left(e^{i \omega t} + e^{-i\omega t}\right) e^{i k_l z +i\Omega_{\rm rot} t \left(J_r-J_0\right)}  \\
\Omega^* \left(e^{i\omega t} + e^{-i\omega t}\right) e^{-i k_l z -i\Omega_{\rm rot} t \left(J_r-J_0\right)} & E_0 - J_0 \Omega_{\rm rot}
\end{pmatrix}
\end{multline}
where $R(t)=e^{i\Omega_{\rm rot} t (J_z + L_z)}$, with $L_z$ the orbital angular momentum of the atom center of mass and $J_z$ the total inner angular momentum of the atom. $\omega$ is the laser frequency, $J_{r/0}$ are the eigenvalues of $J_z$ of the Rydberg and ground states, $E_{R/0}$ are the energies of the Rydberg and ground states and $r$ is the radial coordinate. For the states we considered $J_r=J_0$, but this is not a necessary requirement to obtain the effective Hamiltonian, and it can be relaxed for different trapping techniques. The kinetic part of the Hamiltonian can be recast in the form of a particle in the artificial gauge potential $\vec{A}$. We apply a rotating--wave approximation and we obtain
\begin{equation} \label{hamsp2}
H_{\rm RF} 
= \left[\frac{1}{2m}(\vec{p} + \vec{A})^2 +\frac{m}{2}(\Omega_{tr}^2 - \Omega_{\rm rot}^2) r^2 \right] \mathbb{I} + \begin{pmatrix} 2\delta   & \Omega e^{i k_l z}  \\
\Omega^* e^{-i k_l z} & 0 \end{pmatrix}
\end{equation}
with the detuning $2\delta=E_R-E_0-\omega$. Indeed, the unitary mapping $U_{\rm RW}(t)= e^{i\left[\omega t \frac{(\sigma_z+1)}{2}-\Omega_{\rm rot} t J_z +E_0t\right]}$, needed to apply the rotating--wave approximation, completely erases the effect of the physical rotation $e^{iJ_z \Omega_{\rm rot} t}$ due to the inner degrees of freedom. The rotation of the spin has effects only beyond the rotating--wave approximation in the off--resonant term. The matrix in the right--hand--side of Eq. \eqref{hamsp2} defines indeed the Rydberg dressing that we adopted to obtain the state $\ket{\tilde{g}}$. In this way, within the rotating--wave approximation approximation, we can effectively consider dressed atoms in the state $\ket{\tilde{g}}$ whose dynamics is correctly described by the effective vector potential $\vec{A}$ and by the van der Waals interaction.

	Based on this effective single-particle Hamiltonian, and by considering the long--range interaction in the spirit of \cite{lewenstein2000}, we can write the Gross--Pitaevskii energy in the following form
	\begin{multline}
E=	\int d\vec{r}\, \psi_s^\dag(\vec{r}) \frac{\left(-i\hbar\vec{\nabla} + \vec{A}\right)^2}{2m} \psi_s(\vec{r}) 
+\frac{m\Omega_{\rm tr}^2(1-\gamma^2) r^2}{2} \left|\psi_s(\vec{r})\right|^2 +\\
\int d\vec{r}d\vec{r}' \left|\psi_s(\vec{r}')\right|^2 \left[\frac{g}{2} \delta(\vec{r} - \vec{r}') + \frac{V_6}{a^6 + \left|\vec{r}-\vec{r}'\right|^6}\right]\left|\psi_s(\vec{r})\right|^2 
\end{multline}
In the following, we show that the main effect of the vortex lattice modulation $p$ introduced in $\psi_s$ is to cancel the contribution of the artificial gauge potential $\vec{A}$ to the kinetic energy. Therefore the kinetic energy can be approximated by the one obtained without the gauge potential based on a non-modulated Gaussian wavefunction, thus giving the first term in Eq. 5 of the main text. This will be shown based on a suitable Chern--Simons transformation \cite{zhang92}. The integral for the long--range interaction energy in Eq. 5 of the main text, instead, is obtained by considering the relative coordinate $\vec{r} - \vec{r}'$ and integrating over its orientation.

Let us focus on the kinetic energy term in $H_{RF}$. As we mentioned in the main text, our ansatz for the many--body wavefunction corresponds to $\psi_s(\vec{r})=p(\vec{r})\psi_0(r)$, where $\psi_0(r)=e^{-\frac{r^2}{2s^2}}\sqrt{N/\pi s^2}$ is the normalized Gaussian envelop. We adopted a Gaussian profile for simplicity: alternative approaches based on the interpolation between Gaussian and Thomas--Fermi profiles \cite{komineas2004,watanabe2004} would provide analogous results, with the effect of resulting in a slightly lower average density. Therefore, for the purpose of estimating the onset of the QH regime, the Gaussian ansatz gives more restrictive estimates. The function $p(\vec{r})$ defines instead a hexagonal unit cell with area $2\pi l_B^2$. The average value of its norm is 1, such that the average density of the system is approximately set only by its long--distance behavior, thus by $\psi_0$. Since we are mostly interested in the behavior in the LLL regime, we assume the following analytical form for the periodic function $p$ 
	\begin{equation} \label{pdef}
	p(z) = \frac{\prod_{v \in {\rm V.L.}} \left(z -\eta_v\right)}{\mathcal{N}}\,,
	\end{equation}
where the coordinate $z=x+iy=re^{i\phi}$ and $\eta_v$ is the complex coordinate $x_v + iy_v$ of the vortex $v$ belonging to the vortex lattice. The sum is taken over all the vortices in the lattice. The normalization factor $\mathcal{N}$ is chosen such that
\begin{equation} \label{paverage}
\frac{1}{\mathcal{A}}\int_{\rm u.c.} d\vec{r} \, |p(\vec{r})|^2=1\,,
\end{equation}
where $\mathcal{A} = 2\pi\hbar/B$ is the area of the unit cell of the triangular vortex lattice in the LLL regime.

To show that the main effect of the vortex lattice modulation $p(z)$ is to cancel the artificial gauge potential in the kinetic energy term of the Gross--Pitaevskii equation we apply the following Chern--Simons transformation (see, for example, the review \cite{simon98}, and Ref. \cite{vortexCS} for the application of the Chern--Simons transformation to vortex systems):
\begin{equation}
\psi'(z) = e^{-i\sum_{v\in{\rm V.L.}} \arg\left(z-\eta_v\right)} \psi(z)\,.
\end{equation}
The Chern--Simons phase is the inverse of the phase of $p(z)$, such that $\psi'$ is a real-valued function. This transformation is just a phase change that leaves the density $|\psi|^2$ invariant, and must be treated as a gauge transformation. For the transformed wavefunction $\psi'$ the kinetic energy term reads:
\begin{equation} \label{ekinaa}
E_{\rm kin} = \frac{1}{2m} \int d\vec{r} \, \psi^{\prime\dag}(\vec{r}) \left(-i\hbar \vec{\nabla} + \vec{A} -\vec{a}\right)^2 \psi'(\vec{r})\,,
\end{equation}
where we introduced the Chern--Simons potential:
\begin{equation} \label{aCS}
\vec{a} = -\hbar\vec{\nabla}\left[{\sum_{v\in{\rm V.L.}} \arg\left(z-\eta_v\right)}\right] = \hbar\sum_{v\in{\rm V.L.}} \frac{\hat{x}\left(y-y_v\right)-\hat{y}\left(x-x_v\right) }{\left(x-x_v\right)^2 + \left(y-y_v\right)^2}\,.
\end{equation}
This potential corresponds to a magnetic field $b$ which vanishes everywhere, except at the positions of the vortices
\begin{equation}
\vec{b}(\vec{r})= \vec{\nabla} \times \vec{a} = 2\pi\hbar \hat{z} \sum_{v\in{\rm V.L.}} \delta\left(\vec{r} - \vec{r}_v\right)\,,
\end{equation} 
where $\vec{r}_v=(x_v,y_v)$ is the position of the vortex $v$.
The average value of the amplitude of the field $\vec{b}$ is thus given by the density of the vortices
\begin{equation}
\bar{b} = 2\pi \hbar B/(2\pi \hbar) = B\,.
\end{equation}
We conclude that, on average, the contribution of the phase of the vortices cancels the gauge potential $\vec{A}$, corresponding to the fact that each vortex carries a quantum of angular momentum. Therefore we approximate the kinetic energy of the system by
\begin{equation} \label{ekin}
E_{\rm kin}   \approx -\frac{\hbar^2}{2m} \int d\vec{r} \, \psi^{\prime \dag}\vec{\nabla}^2\psi' =  \frac{\hbar^2}{2m} \int d\vec{r} \left[\left|p(\vec{r})\right|^2 (\vec{\nabla}\psi_0)^2 -\psi_0^2 |p(\vec{r})|\vec{\nabla}^2|p(\vec{r})|\right] \,.
\end{equation}
To evaluate this expression, we apply the so-called averaged vortex approximation \cite{ho2001}: for a system in a strong magnetic field such that $s \gg l_B$ and the area of the unit cell of the vortex lattice is much smaller than the system size, we can separate the rapidly oscillating contributions proportional to $p$ from the global Gaussian contribution of $\psi_0$.
In particular we consider that $|p|^2$ averages to 1 whereas $-|p|\vec{\nabla}^2|p|$ averages to a constant $c$ which depends on $B$ only. We finally obtain:
\begin{equation}
E_{\rm kin}   \approx \frac{\hbar^2}{2m} \int d\vec{r} \left[(\vec{\nabla}\psi_0)^2 +\psi_0^2 c\right] =  \frac{\hbar^2 N}{2ms^2} + c\,.
\end{equation}
The constant $c$ does not depend on the parameter $s$; therefore it can be dropped in Eq. \eqref{GP} because it has no effect in the estimate \eqref{effint}.

A similar separation between slowly and rapidly varying contributions applies to the estimate of the trapping energy (the second term in Eq. \eqref{GP}). In this case, the rapidly oscillating modulation $|p|^2$ averages to 1, thus leaving only the result related to the Gaussian envelope. Concerning the contact interaction, instead, the role of the modulation is to increase the effective interaction $g$ to $bg$ with $b= \int_{\rm u.c.} d\vec{r} \left|p(\vec{r})\right|^4 \approx 1.1596$. This holds in the LLL approximation, whereas for weaker magnetic fields $B$, thus smaller densities of vortices, the factor $b$ decreases to $1$ in the limit $B\to 0$.

The averaged vortex approximation allows us also to show that the parameter $\alpha_\xi$ does not depend on the long-range Rydberg interactions. We observe indeed that$\alpha_\xi$ is related to the ratio $2\alpha_\xi^2$ between the vortex core and the unit cell area at the cross-over between the superfluid and LLL regimes. In the superfluid phase, $p$ does not match the analytic function \eqref{pdef}. Its profile $|p|$ and the size of the vortex core can be determined by minimizing the kinetic and interaction energy of the superfluid in the unit cell \cite{baym04}. In the limit $s\gg l_B$, the core area is essentially independent of the long-range Rydberg interactions since its contribution to the energy is independent of $p$. This is because of two reasons: (i) The leading contribution of the interaction energy is provided by the product of densities in well-separated unit cells, such that these densities, effectively, average to the value provided by the slowly varying Gaussian envelope and are not affected by $p$. (ii) The residual contribution of the density-density interaction within the same cell is independent of $p$ because the interaction profile is flat at short distances and the Rydberg radius $a$ in our regime of interest is typically larger than $l_B$; thus this contribution results in:
 \begin{equation}
\int_{\rm u.c.} d\vec{r}\, d\vec{r}' |p(\vec{r})|^2 |p(\vec{r}')|^2 V_{\rm Rydberg} (\vec{r}-\vec{r}') \approx V_6 \mathcal{A}^2/a^6\,,
\end{equation}
which is independent of $p$. Therefore, we conclude that only the short range contact interactions influence the profile $p$, and the value $\alpha_\xi$ is not affected by the long-range interactions.

\section{Scattering length of Rydberg-dressed atoms} \label{app:scatter}

We estimate here the scattering length determined by the van der Waals effective interaction.
Our starting point is the Lippmann--Schwinger equation for the scattering state $\chi_{\vec{k}}(\vec{r})$. For the scattering in a 3D system with isotropic interactions,
\begin{equation}
\chi_{\vec{k}}(\vec{r}) = \chi_{0,k}(\vec{r}) - \frac{m_r}{2\pi \hbar^2} \int d^3 \vec{r}' \frac{e^{ik\left|\vec{r}-\vec{r}'\right|}}{\left|\vec{r}-\vec{r}'\right|} \frac{V_6}{a^6 + \left|r'\right|^6}\chi_{\vec{k}}(\vec{r}')\,,
\end{equation}
where $\chi_{0,\vec{k}}(\vec{r})=e^{i\vec{k}\vec{r}}$ is an incoming plane wave, $m_r=m/2$ is the reduced mass and we use the shorthand notation $v=\left|\vec{v}\right|$ for the moduli of momentum and position vectors. In particular we express the scattering wavefunction $\chi_{\vec{k}}(\vec{r})$ as
\begin{equation}
\chi_{\vec{k}}(\vec{r}) = e^{i\vec{k}\vec{r}} + \frac{e^{ikr}}{r}f(\vec{k}',\vec{k})\,,
\end{equation}
where $\vec{k'}=k \hat{r}$ is the outgoing wave-vector and $f$ is the scattering amplitude. From the previous equations we derive
\begin{equation} \label{amplitude}
f(\vec{k}',\vec{k})\left[-\frac{2\pi\hbar^2}{m_r V_6}+\frac{2\pi}{ik} \int_0^\infty dr' \frac{1-e^{2ikr'}}{a^6 + r^{\prime 6}}\right]=2\pi \int_0^\infty dr'\frac{2r'}{a^6+ r^{\prime 6}}\frac{2\sin\left(r'\left|\vec{k}-\vec{k}'\right|\right)}{\left|\vec{k}-\vec{k}'\right|}\,.
\end{equation}
We consider only the isotropic s-wave component of the scattering amplitude by taking the angular average of the previous equation. In particular, we define
\begin{equation}
f_0 = \frac{1}{4\pi} \int d\Omega_r f(\vec{k}',\vec{k})\,,
\end{equation}
where $\Omega_r$ is the direction of the vectors $\vec{r}$ and $\vec{k}'$. By integrating Eq. \eqref{amplitude} over $\Omega_r$ we obtain
\begin{equation}
f_0 = \frac{-2\pi}{3a^3}\left(\frac{\hbar^2}{m_rV_6} + \frac{2\pi}{3\sqrt{3}a^4}\right)^{-1}.
\end{equation}
Considering the s-wave component of the scattering matrix $S_0 = 1+2ikf_0$ and taking the limit $k \to 0$ we obtain the scattering length:
\begin{equation} \label{scattering}
a_s = -f_0= \frac{2\pi}{3a^3}\left(\frac{\hbar^2}{m_rV_6} + \frac{2\pi}{3\sqrt{3}a^4}\right)^{-1}\,.
\end{equation}
We observe that in the Born approximation the last term in the parenthesis would be neglected and the result matches the calculation in Ref. \cite{buechler2010}. Eq. \eqref{scattering} results in a scattering length of approximately $0.83$ \textmu m for the typical values of the Rydberg dressing described in the main text, in proximity to the optimal point, thus to $\theta \approx 0.05$.


\begin{thebibliography}{50}

\bibitem{bloch08} I. Bloch, J. Dalibard, and W. Zwerger,
  Rev. Mod. Phys. \textbf{80}, 885 (2008).

\bibitem{dalfovo99} F. Dalfovo, S. Giorgini, L. P. Pitaevskii  and S. Stringari
Rev. Mod. Phys. \textbf{71}, 463 (1999).
  

\bibitem{cornell04} V. Schweikhard, I. Coddington, P. Engels, V.P. Mogendorff and E.A. Cornell, Phys. Rev.
Lett. \textbf{92},  040404 (2004).
\bibitem{dalibard04} V. Bretin, S. Stock, Y. Seurin and J. Dalibard, Phys. Rev. Lett. \textbf{92}, 050403 (2004).
\bibitem{fetter09} A. L. Fetter, Rev. Mod. Phys. \textbf{81}, 647 (2009).

\bibitem{baym05} G. Baym, J. Low. Temp. Phys., \textbf{138}, 601 (2005).
\bibitem{cooper08} N. R. Cooper, Adv. Phys. {\bf 57}, 539 (2008).

\bibitem{spielman09} Y.-J. Lin, R. L. Compton, K. Jim\'enez-Garc\'ia, J. V. Porto, and I. B. Spielman, Nature \textbf{462}, 628-632 (2009). 
\bibitem{spielman13} M. C. Beeler, R. A. Williams, K. Jiménez-García, L. J. LeBlanc, A. R. Perry and I. B. Spielman, Nature \textbf{498}, 201-204 (2013)
\bibitem{dalibard11}
J. Dalibard, F. Gerbier, G. Juzeli\={u}nas and P. \"Ohberg
Rev. Mod. Phys. \textbf{83}, 1523 (2011). 
\bibitem{goldman13rev} N. Goldman, G. Juzeli\=unas, P. \"Ohberg and I. B. Spielman,  Rep. Prog. Phys. \textbf{77}, 126401 (2014).

\bibitem{modugno2019}  L.  Tanzi,  E.  Lucioni,  F.  Fam\`a,  J.  Catani,  A.  Fioretti, C. Gabbanini, R. N. Bisset, L. Santos  and G. Modugno, Phys. Rev. Lett. {\bf 122}, 130405 (2019).
\bibitem{pfau2019} F.  B\"ottcher,  J.-N.  Schmidt,  M.  Wenzel,  J.  Hertkorn, M. Guo, T. Langen  and T. Pfau, Phys. Rev. X \textbf{9}, 011051 (2019).
\bibitem{ferlaino2019}  L. Chomaz, D. Petter, P. Ilzh\"ofer, G. Natale, A. Trautmann,  C.  Politi,  G.  Durastante,  R.  M.  W.  van  Bijnen, A.  Patscheider,  M.  Sohmen,  M.  J.  Mark and  F.  Ferlaino, Phys. Rev. X {\bf 9}, 021012 (2019).
\bibitem{ferlaino2019b} G. Natale, R. M. W. van Bijnen, A. Patscheider, D. Petter, M. J. Mark, L. Chomaz, and F. Ferlaino
Phys. Rev. Lett. \textbf{123}, 050402 (2019).

\bibitem{gross15} J. Zeiher, P. Schauss, S. Hild, T. Macr\`i, I. Bloch and C. Gross, Phys. Rev. X {\bf 5}, 031015 (2015).
\bibitem{gross16} J. Zeiher, R. van Bijnen, P. Schauss, S. Hild, J.-Y. Choi, T. Pohl, I. Bloch and C. Gross, Nat. Phys. {\bf 12}, 1095 (2016).
\bibitem{bernien17} H. Bernien, S. Schwartz, A. Keesling {\it et al.}, Nature \textbf{551}, 579 (2017).

\bibitem{Grusdt2013} F. Grusdt and M. Fleischhauer, Phys. Rev. A \textbf{87}, 043628 (2013). 
\bibitem{Grass2018} T. Grass, P. Bienias, M. J. Gullans, R. Lundgren, J. Maciejko and A. V. Gorshkov, Phys. Rev. Lett. \textbf{121}, 253403 (2018).

\bibitem{noteb} For neutral atoms we choose units for $B$ given by mass/time, such that the magnetic flux has the units of an angular momentum.


\bibitem{rozhkov96} A. Rozhkov and D. Stroud, Phys. Rev. B \textbf{54}, R12697(R) (1996)

\bibitem{sinova02} J. Sinova, C. B. Hanna and A. H. MacDonald, Phys. Rev. Lett. \textbf{89}, 030403 (2002).

\bibitem{cooper01} N.R. Cooper, N.K. Wilkin and J.M.F. Gunn, Phys. Rev. Lett. \textbf{87},  120405 (2001).
\bibitem{regnault2004} N. Regnault and T. Jolicoeur, Phys. Rev. B \textbf{69}, 235309 (2004).


\bibitem{gemelke10} N. Gemelke, E. Sarajlic and S. Chu, arXiv:1007.2677.



\bibitem{low2012} R. L\"ow, H. Weimer, J. Nipper, J. B. Balewski, B. Butscher, H. P. B\"uchler and T. Pfau, J. Phys. B: At. Mol. Opt. Phys. \textbf{45}, 113001 (2012).


\bibitem{lewenstein2000} L. Santos, G. V. Shlyapnikov, P. Zoller and M. Lewenstein, Phys. Rev. Lett. \textbf{85}, 1791 (2000).


\bibitem{aftalion2005} A. Aftalion, X. Blanc and J. Dalibard, Phys. Rev. A \textbf{71}, 023611 (2005).

\bibitem{ho2001} T.-L. Ho, Phys. Rev. Lett. \textbf{87}, 060403 (2001).

\bibitem{coddington04} I. Coddington, P. C. Haljan, P. Engels, V. Schweikhard, S. Tung, and E. A. Cornell, Phys. Rev. A \textbf{70}, 063607 (2004).
\bibitem{baym04} G. Baym and C. J. Pethick, Phys. Rev. A \textbf{69}, 043619 (2004).
\bibitem{baymnote} $2\alpha_\xi^2 = 0.173 - 0.225$ is the fractional core area of the vortices estimated in \cite{baym04} and measured in \cite{cornell04,coddington04} at the crossover between the normal superfluid and LLL regimes. Its value, as discussed in \cite{baym04}, relies mostly on the short distance behavior of the interaction and is not influenced by the long-range Rydberg interaction (see Appendix \ref{app:rotating}).

\bibitem{Pohl2012} N. Henkel, F. Cinti, P. Jain, G. Pupillo and T. Pohl, Phys. Rev. Lett. \textbf{108}, 265301 (2012).


\bibitem{Haldane} F. D. M. Haldane, Phys. Rev. Lett. \textbf{51}, 605 (1983).


\bibitem{cooper2005} N. R. Cooper, E. H. Rezayi and S. H. Simon, Phys. Rev. Lett. \textbf{95}, 200402 (2005).
\bibitem{zhang2005} J. Zhang and H. Zhai, Phys. Rev. Lett. \textbf{95}, 200403 (2005).




\bibitem{Henckel2010} N. Henkel, R. Nath and T. Pohl, Phys. Rev. Lett. \textbf{104}, 195302 (2010).
\bibitem{igor10} G. Pupillo, A. Micheli, M. Boninsegni, I. Lesanovsky and P. Zoller, Phys. Rev. Lett. \textbf{104}, 223002 (2010).
\bibitem{Macri2014} F. Cinti, T. Macr\'i, W. Lechner, G. Pupillo and T. Pohl, Nature Comm. \textbf{5}, 3235 (2014).
\bibitem{Macri2013}T. Macr\'i, F. Maucher, F. Cinti and T. Pohl, Phys. Rev. A \textbf{87}, 061602(R) (2013).
\bibitem{Sinha2005} S. Sinha and G. V. Shlyapnikov, Phys. Rev. Lett. \textbf{94}, 150401 (2005).

\bibitem{giovanazzi09} S. Giovanazzi, J. Esteve and M. K. Oberthaler, New J. Phys. \textbf{10} 045009 (2009).


\bibitem{cornell2003} P. Engels, I. Coddington, P. C. Haljan, V. Schweikhard, and E. A. Cornell, Phys. Rev. Lett. \textbf{90}, 170405 (2003).
\bibitem{juze06} G. Juzeli\={u}nas, J. Ruseckas, P. \"Ohberg and M. Fleischhauer, Phys. Rev. A \textbf{73}, 025602 (2006).

\bibitem{roncaglia11} M. Roncaglia, M. Rizzi and J. Dalibard, Sci. Rep. {\bf 1}, 43 (2011).
\bibitem{ho16} Tin-Lun Ho, arXiv:1608.00074 (2016).



\bibitem{heidemann07} R. Heidemann, U. Raitzsch, V. Bendkowsky, B. Butscher, R. L\"ow, L. Santos and T. Pfau,
Phys. Rev. Lett. \textbf{99}, 163601 (2007).


\bibitem{cornell96} D. S. Jin, J. R. Ensher, M. R. Matthews, C. E. Wieman and E. A. Cornell, Phys. Rev. Lett. \textbf{77}, 420 (1996). 

\bibitem{bounds2018}A. D. Bounds, N. C. Jackson, R. K. Hanley, R. Faoro, E. M. Bridge, P. Huillery and M. P. A. Jones,
Phys. Rev. Lett. \textbf{120}, 183401 (2018).

\bibitem{zhang92} S.-C. Zhang, Int. J. Mod. Phys. \textbf{6}, 25 (1992).

\bibitem{komineas2004}  N. R. Cooper, S. Komineas and N. Read, Phys. Rev. A \textbf{70}, 033604 (2004).
\bibitem{watanabe2004} G. Watanabe, G. Baym and C. J. Pethick, Phys. Rev. Lett. \textbf{93}, 190401 (2004).


\bibitem{simon98} S. H. Simon, ``THE CHERN--SIMONS FERMI LIQUID DESCRIPTION OF FRACTIONAL QUANTUM HALL STATES'', pp. 91-194 in ``Composite Fermions'', ed. O Heinonen, World Scientific (1998).
\bibitem{vortexCS}  S. V. Iordanski and D. .S Lyubshin, J. Phys.: Condens. Matter \textbf{21}, 405601 (2009). 

\bibitem{buechler2010} J. Honer, H. Weimer, T. Pfau and H. P. B\"uchler, Phys. Rev. Lett. \textbf{105}, 160404 (2010).




\end{thebibliography}
\end{document}